\documentclass[usenatbib]{mn2e}
\input{epsf}

\def\etal{et al.}

\def\alm{a_{\ell m}}
\def\Ylm{Y_{\ell m}}
\def\Cl{C_{\ell}}

\def\summ{\sum_{m=-\ell}^{\ell}}
\def\suml{\sum_{\ell=0}^{\infty}}
\def\a{{\bf a}}

\def\Ph{{\bf \Psi}}

\def\Ks{{\bf \Xi}}
\def\ks{{\bf \xi}}
\def\G{{\bf G}}

\def\w{{\bf w}}

\newcommand{\nbi}{{Niels Bohr Institute, Blegdamsvej 17,
DK-2100 Copenhagen, Denmark}}
\newcommand{\sao}{{Special Astrophysical Observatory, Nizhnij Arkhyz,
Karachaj-Cherkesia, 369167, Russia}}

\newcommand{\ru}{{Rostov State University, Space Research Department,
Zorge,5, 344091, Russia}}

\title[Cross-correlation of the CMB and foregrounds phases]
{Cross-correlation of the CMB and foregrounds phases derived 
from the {\it WMAP} data}

\author[Naselsky, Doroshkevich \& Verkhodanov]
{ P. D. Naselsky$^{1-3}$, A.Doroshkevich$^{1}$, O. 
Verkhodanov$^{1,4}$\\
$1$Theoretical Astrophysics Center, Juliane Maries Vej
30, DK-2100,  Copenhagen, Denmark.\\
$2${\nbi}\\
$3${\ru}\\
$4${\sao}
}

\date{Accepted ...,
      Received ...,
      in original form ... .}

\begin{document}
\maketitle

\begin{abstract}
We present circular and linear  cross-correlation tests and
the ``friend--of--friend'' analysis for phases
of the Internal Linear Combination Map (ILC)  and the {\it WMAP} foregrounds  for all K--W frequency bands
 at the range of multipoles $\ell\le100$.
We compare also Tegmark, de Oliveira--Costa and Hamilton (2003)
and Naselsky et al. (2003) cleaned maps with corresponding
foregrounds.
We have found significant deviations from the expected Poissonian
statistics for all the cleaned maps and foregrounds.
Our analysis shows that, for a low multipole range of the cleaned 
maps, power spectra contains some of the foregrounds residuals 
mainly from the W band.

\end{abstract}
\begin{keywords}cosmology: cosmic microwave background --- cosmology:
observations --- methods: data analysis
\end{keywords}

\section{Introduction}

The recently-published Wilkinson Microwave Anisotropy Probe
({\it WMAP}) data sets (see Bennett et al. 2003 a-c, Hinshaw 
et al. 2003 a-b) open a new epoch for the CMB investigation.
These data serve
for the development of more refined technique for future high
resolution measurements and for a choice of the realistic 
cosmological model.

The {\it WMAP} has observed the full sky in five frequency
bands: K~ (centered frequency 22.8 GHz), Ka~(33.0 GHz),Q~(40.7 
GHz), V~(60.8 GHz), and W~(93.5 GHz) and produced five maps
represented in $n=12\times 512^2$ HEALPix (G\'orski et al. 1999)
pixels. Then, using smoothing of the maps to $1^o$ and performing 
cleaning of the combined multifrequency map by minimization of 
the rms variance for each pixel, the {\it WMAP} team produced 
the Internal Linear Combination (ILC) map taking 12 optimization
coefficients $w_i, i=1,2,...,5$ for 12 disjoint sky regions. Thus,
the ILC map has a minimal variance corresponding to the minimization of
the Galaxy and foreground contamination at the range of multipoles 
$\ell\le 100$.

\begin{figure*}
\begin{minipage}{160mm}
\centering
\hbox{\hspace{0.9cm}\epsfxsize=7.cm
\epsfbox{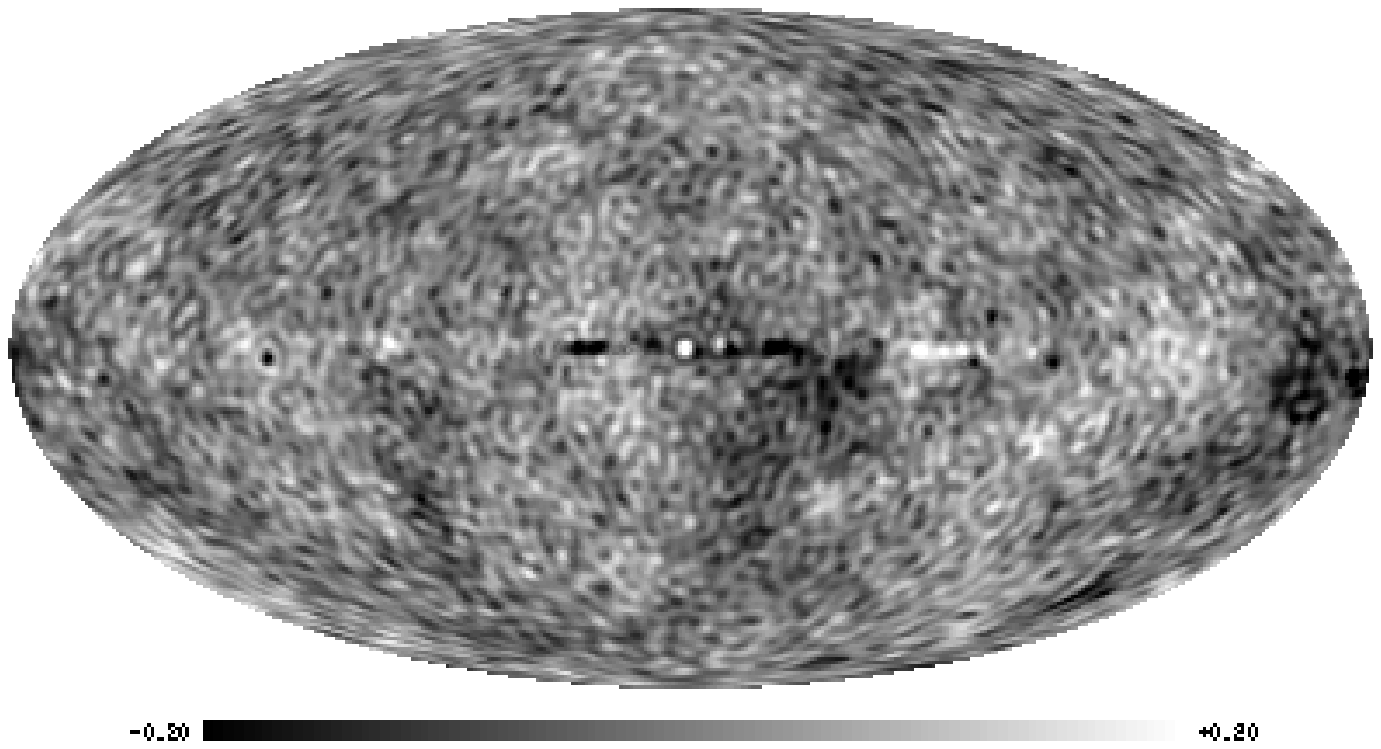}\epsfxsize=7.cm
\epsfbox{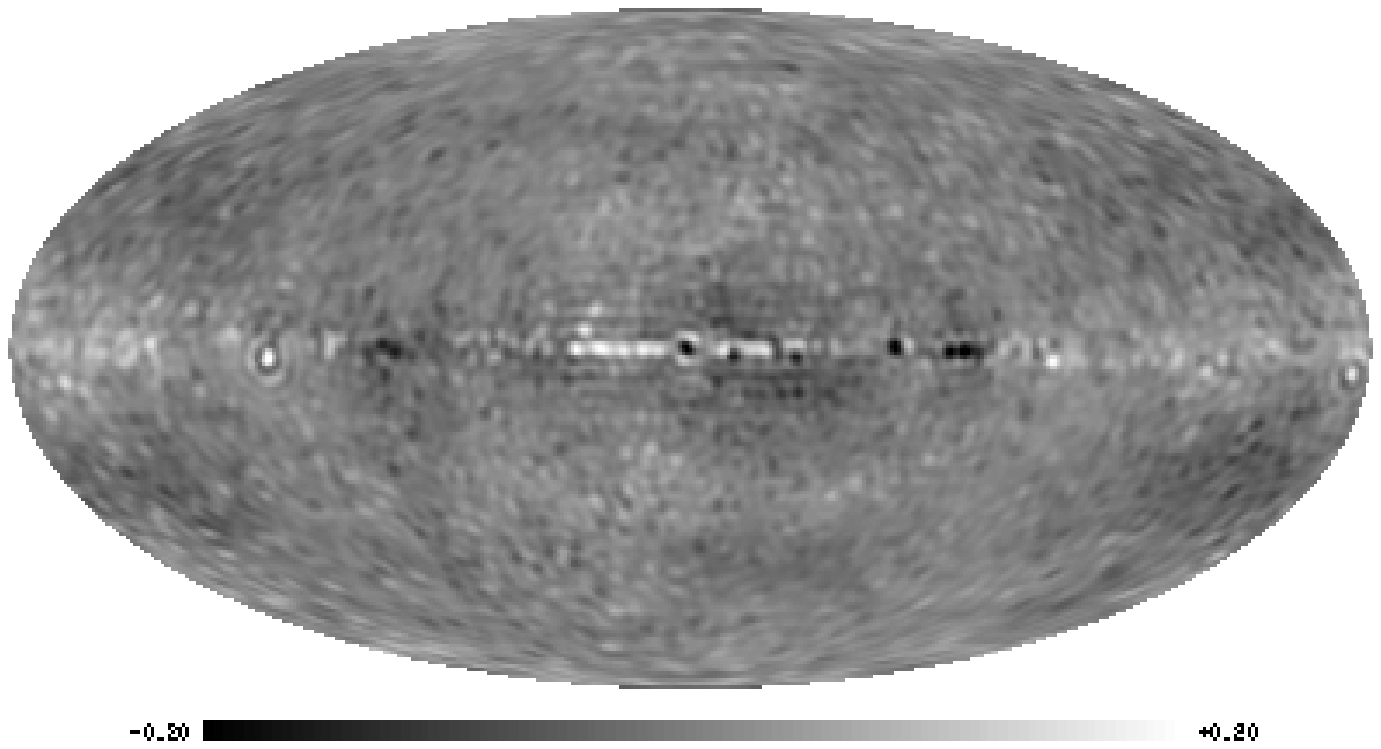}}
\hbox{\hspace*{0.9cm}\epsfxsize=7.cm
\epsfbox{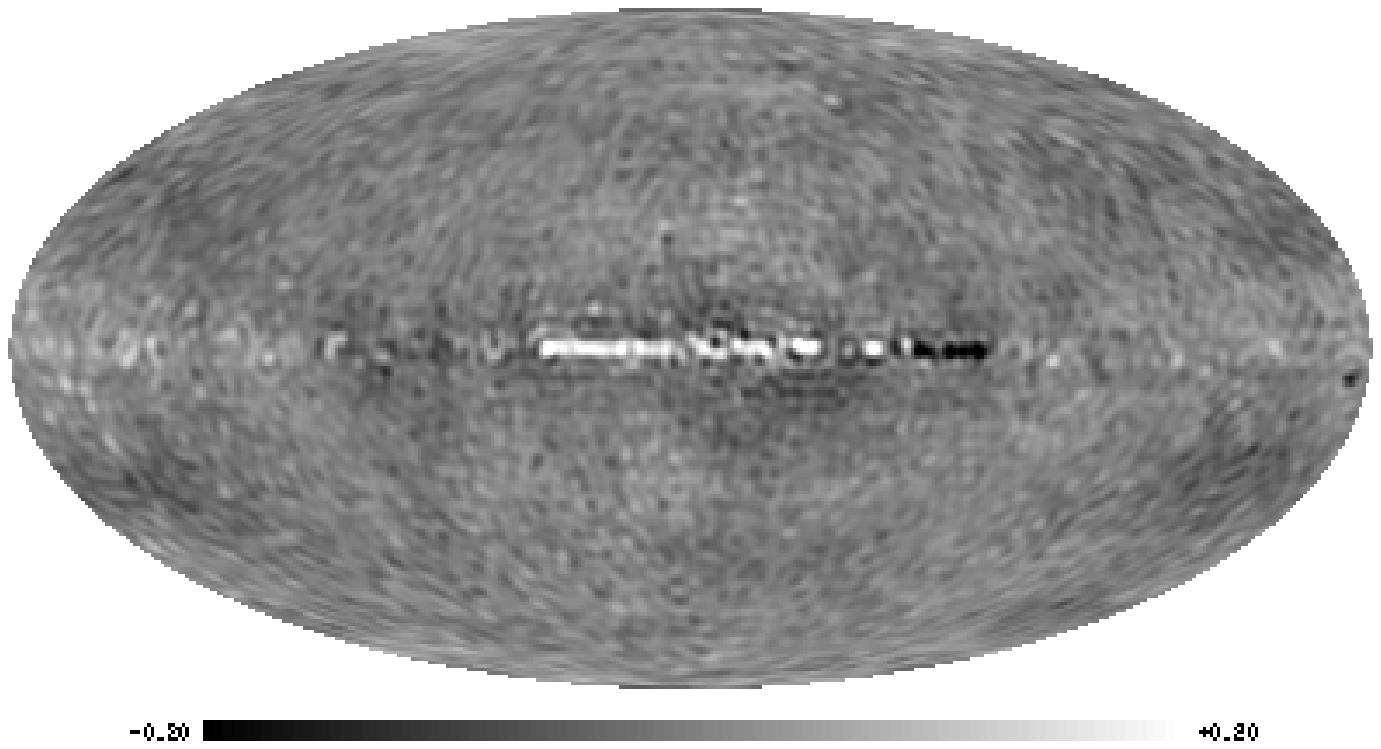}\epsfxsize=7.cm
\epsfbox{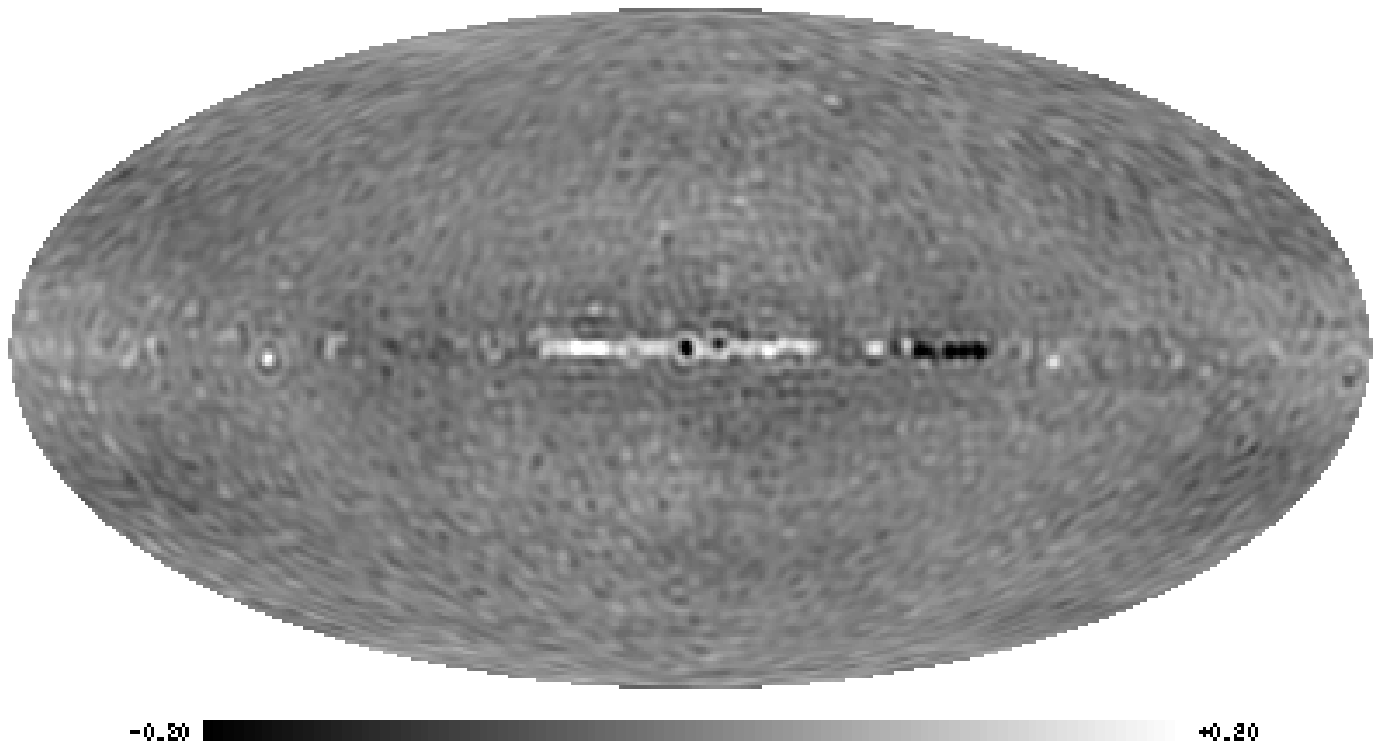}}
\vspace{1.cm}
\caption{The map for reconstructed CMB at the range of multipoles 
$\ell\le 100$ (top left) and differences between ILC and PCM maps
(top right), TOH FCM and PCM (bottom left) and
TOH Wiener and PCM maps (bottom right).
} 
\label{f1}
\end{minipage}
\end{figure*}

Tegmark, de Oliveira-Costa and Hamilton (2003)
(hereafter TOH) have suggested to use ideologically similar but 
technically different 
optimization scheme based on Tegmark and Efstathiou (1996) 
(hereafter TE) method. The main goal of the TOH method is to combine 
the K--W band maps into the Foreground Cleaned Map (FCM) using
variable weighting coefficients $w_i(\ell)$ for each frequency band
$i$=K--W.  In order to minimize the total unweighted power from
the Galaxy, foregrounds and noise separately for each harmonic 
$\ell$,  TOH subdivided the sky into 9 disjoint regions and performed
cleaning of the 5 frequency  maps in each region. Unlike the ILC map,
the TOH FCM has smaller power at the multipole range $\ell\le 12$.
Both the ILC and TOH FCM are available from the web and can be used for
analysis of the foregrounds and the statistical properties of the cleaned
maps. For example, Komatsu et al. (2003) tested the non--Gaussianity
of the {\it WMAP} CMB signal using Minkowski functionals as a 
statistic. Chiang et al. (2003) have also tested the statistics of
the phases for the ILC and the TOH FCM maps and discussed non-Gaussianity of
the TOH FCM map. Dineen and Coles (2003) have presented a diagnostic 
of the Galactic synchrotron contamination and have discovered the
cross--correlation between Faraday rotation measure and the TOH FCM map
(see also Coles et al. 2003).
  
Recently Naselsky et al. (2003) proposed the Phase Cleaning Method (PCM) 
for decomposition of the CMB signal and foregrounds using the {\it WMAP}
data. The main idea of the PCM is to minimize the variance of the 
combined K--W maps and cross--correlations of the CMB and foreground  
phases simultaneously. The result of PCM in its application to the
{\it WMAP} data reproduces well the power spectrum of the
best-fit {\it WMAP} $\Lambda$CDM model at the range $\ell<50$.
The main target of our paper is to extend the resolution of the 
derived PCM CMB map up to $\ell\sim 100$, to perform the the 
cross-correlation analysis (circular and linear statistics and 
``friend--of--friend'' analysis)
of the CMB and foreground phases and to estimate their interconnection. 
We have included the ILC , TOH FCM and the PCM maps
in our analysis and have
discovered significant deviation of the phase statistics from the expected 
for uniformly random phases.
We have shown that such a kind of non-Gaussianity
arises owing to the contamination of the foregrounds signal
in the ILC, TOH FCM and
the PCM maps.

\section{Phase cleaning method and the CMB extraction}

The fluctuations of measured CMB plus foregrounds radiation
on a sky sphere can be expressed as a sum over spherical harmonics:
\begin{equation}
\Delta T(\theta,\varphi)=T(\theta,\varphi)- \langle T \rangle =
	     \suml \summ \alm \Ylm (\theta,\varphi),
\label{eq1}
\end{equation}
where $\alm $ are the coefficients of expansion,
$\langle T \rangle$=2.73K, and $\langle\Delta T\rangle$=0.
Homogeneous and
isotropic CMB Gaussian random fields (GRFs), as a result of the 
simplest inflation paradigm, possess $\alm$ modes whose real
and imaginary parts are independently distributed. The statistical
properties of this field are completely specified
by its angular power spectrum $\Cl$, 
\begin{equation}
\langle  a^{cmb}_{\ell^{ } m^{ }} (a^{cmb})^{*}_{\ell^{'} m^{'}}
\rangle = \Cl^{cmb} \; \delta_{\ell^{ } \ell^{'}} \delta_{m^{} m^{'}},
\label{eq2}
\end{equation}
and random phases
\begin{equation}
\Psi^{cmb}_{\ell m}=\tan^{-1}\frac{Im (\alm^{cmb})}{Re (\alm^{cmb})},
\label{eq3}
\end{equation}
which are uniformly distributed at the range $0, 2\pi$. For the
foregrounds, the signal is obviously non-Gaussian. So, for the
combined CMB + foregrounds signal  we define
\begin{equation}
\alm=|\alm|\exp(i\Psi_{\ell m}),
\label{eq4}
\end{equation}
where $|\alm|$ is the modulus and $\Psi_{\ell m}$ is the phase of
each $\ell, m$ harmonic. In practice, each of the {\it WMAP} K--W maps is
decomposed into set of $\a^j$ where $\a\equiv \alm$ and index 
$j=1,2..5$ corresponds to K, Ka, ..., W bands, using the HEALPix code
(G\'orski et al. 1999).\footnote{In addition we used the GLESP
pixelization scheme (Doroshkevich et al. 2003a) in order to
control the accuracy of the  $\a^{(j)}$ harmonics estimation.}

The basic idea of the PCM is to generalize the TOH and TE minimization 
scheme including also the minimization of the cross-correlations between 
derived CMB signal and foregrounds (Naselsky et al. 2003). The method
does not require any galactic cut--offs or disjoint regions. 
Following Naselsky et al. (2003), we consider the combinations of 
the {\it WMAP} maps: Ka--Q, Ka--V and Q--V, for which the higher 
correlations between foreground phases takes place.
\footnote{We did not include W and K bands to the separation
procedure because of the peculiar phases (Naselsky et al. 2003).}.
For each pairs of the maps, we introduce weighting coefficients
$\w^{(j)}(\ell)$ similar to TOH and TE methods and minimize the 
variance of the derived map per each mode $\ell$. Neglecting the 
instrumental for $\ell\le 100$ noise and taking into account the 
beam shape deconvolution, we consider the signal as a
superposition of the CMB and the foregrounds.

From all $\a^{(j)}$, we have found the phases
$\Psi^{(j)}_{\ell m}\equiv\Ph^{(j)}$ (see Eq~(\ref{eq2})--(\ref{eq4})). 
Each set of the  $\a^{(j)}$ coefficients
is defined by a combination of  the different foreground coefficients
$\sum_k\a^{(j)}_k=\G^{(j)}$ and the CMB signal $\a^{cmb}$, where index
$k$ marks the synchrotron, free-free and dust emission. For each
combination of the maps, the derived CMB map is defined as
\begin{equation}
a^{M}_{\ell m}=\sum_{j=1}^2  \w^{(j)}(\ell)a^{(j)}_{\ell m}
	     = a^{cmb} + \sum_{j=1}^2  \w^{(j)} G^{(j)},
\label{eq5}
\end{equation}
where 
\begin{eqnarray}
\w^{(1)}(\ell)=
     \frac{\sum_m \left\{ |\a^{(2)}| \left(|\a^{(2)}|-|\a^{(1)}|
       \cos(\Ks^{(2)}-\Ks^{(1)})\right)\right\}}
	      {\sum_m|\a^{(1)}-\a^{(2)}|^2}, &\nonumber\\
\w^{(2)}(\ell)=\frac{\sum_m\left\{|\a^{(1)}| \left(|\a^{(1)}|-|\a^{(2)}|
       \cos(\Ks^{(2)}-\Ks^{(1)}) \right)\right\}}
	      {\sum_m|\a^{(1)}-\a^{(2)}|^2},
\label{eq6}
\end{eqnarray}
and $\Ks^{(j)}$ is the phase of $j$-channel.

The phases of the reconstructed CMB are related to the foregrounds
amplitudes and phases as follows (Naselsky et al. 2003)
\begin{equation}
\Ph^{M}= \ks + \arcsin\frac{\sum_j \w^{(j)}|\G^{(j)}|\sin({\Ph^{(j)}-\ks)}
\cos(\Ph^{(M)})}{\sum_j \w^{(j)}|\G^{(j)}|\cos{\Ph^{(j)}} + |\a^{cmb}|
\cos{\ks}},
\label{eq7}
\end{equation}
where $\ks$ is the true CMB phase and $\Ph^{(j)}$ are the foreground
phases. As is mentioned in Naselsky et al. (2003), the phases of
the reconstructed CMB signal $\Ph^{M}$ would have correlations
with those of foregrounds. The main goal of the PCM is to
minimize such correlations by minimizing of the weighting variance
\begin{equation}
V=\frac{1}{2\pi(2\ell+1)}\sum_m\frac{|\a^{cmb}|^2}{C_{\ell}}
\int_{0}^{2\pi}d\ks\left(\Ph^{M}-\ks\right)^2\rightarrow min
\label{eq8}
\end{equation}
using Eq.(\ref{eq7}) and weighting coefficients $\w^j$ in a form
\begin{eqnarray}
\w^{(1)}=\frac{\sum_m \left\{|\G^{(2)}|\left[|\G^{(2)}|-|\G^{(1)}|
	     \cos(\Ph^{(2)}-\Ph^{(1)})\right]\right\}}
		  {\sum_m |\G^{(1)}-\G^{(2)}|^2} &\nonumber\\
\w^{(2)}=\frac{\sum_m \left\{|\G^{(1)}|\left[|\G^{(1)}|-|\G^{(2)}|
	     \cos(\Ph^{(2)}-\Ph^{(1)})\right]\right\}}
		  {\sum_m |\G^{(1)}-\G^{(2)}|^2}.
\label{eq9}
\end{eqnarray}
Note that minimization of the weighting variance Eq.(\ref{eq8})
is equivalent to minimization of the error bars of the CMB 
reconstruction $\sum_m|a^{M}_{\ell m}-a^{cmb}_{\ell m}|^2\rightarrow \min$,
where $a^{cmb}_{\ell m}$ is the true CMB (Naselsky et al. 2003).

\begin{figure}
\centering
\epsfxsize=8.cm
\epsfbox{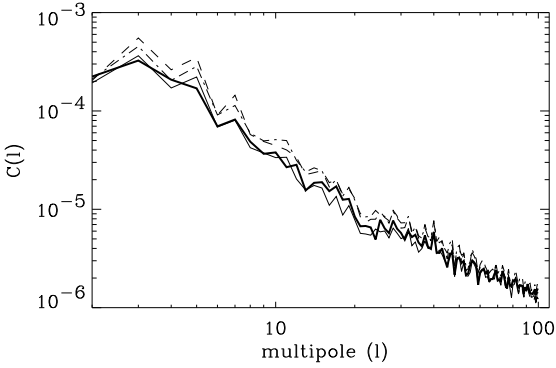}
\caption{The power spectrum for the PCM reconstructed CMB (thick
solid line), ILC (dash line), TOH FCM 
(dash dot line) and TOH Wiener filtered map (thin solid line).
} 
\label{f2}
\end{figure}

Practical implementation of the PCM includes the iteration
scheme in which the first step corresponds to minimization of 
the variance of the derived map in each mode $\ell$ with the choice
of coefficients Eq.(\ref{eq6}). This step of optimization
reconstructs the pre-CMB and pre-foreground map for each pair
of channels where the foreground maps are a simple subtraction
$\alm$ coefficients of the K--W signals and derived pre-CMB
$\alm$. Then, using the moduli and phases of the
pre-foregrounds we perform the next iteration, using Eqs.(\ref{eq9}),
(\ref{eq5}) and so on. This iteration scheme is stable and
reproduce quite well the CMB signal from each pair of K--W 
maps after two steps of iteration. 

\begin{figure}
\centering
\epsfxsize=7.cm
\epsfbox{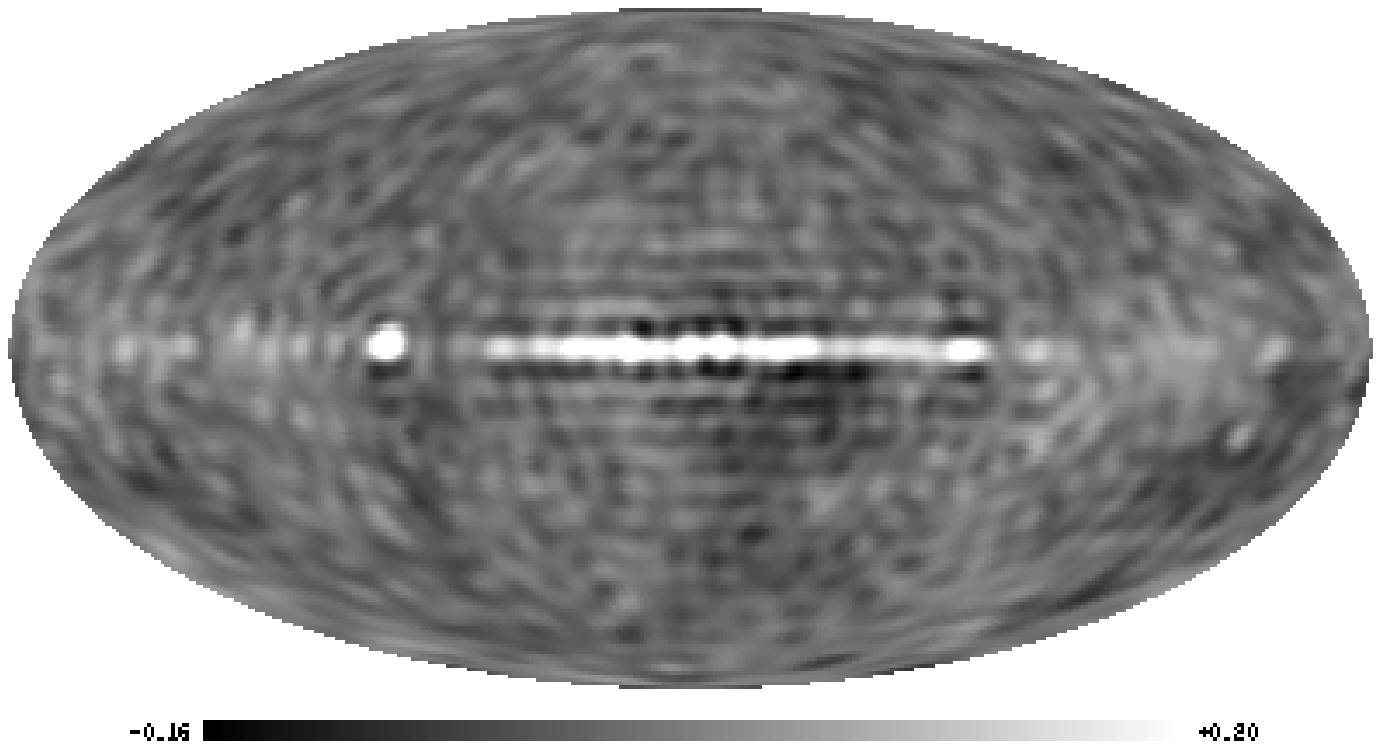}
\vspace{0.5cm}
\epsfxsize=7.cm
\epsfbox{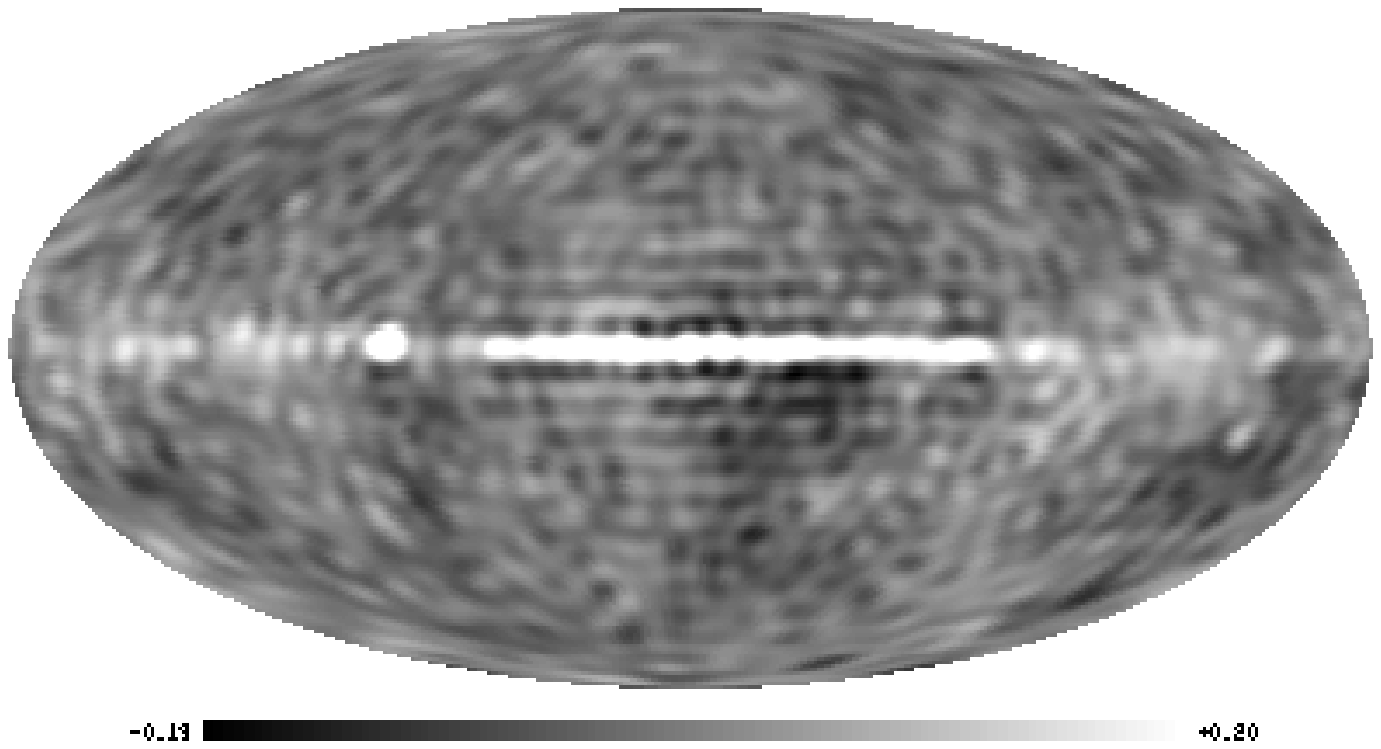}
\vspace{0.6cm}
\caption{The maps for differences $S-ILC-Foregrounds$ for 
V (top) and W (bottom) {\it WMAP} bands.
} 
\label{f3}
\end{figure}

However, the CMB maps reconstructed from each pair of K--W 
maps are slightly different because of residues
of the foregrounds. To minimize such residues, as the last
step of the PCM we use the so called MIN-MAX filter for each pixels
of the map. The MIN-MAX filter compares signals, $\Delta T_p$, 
in each pixel $p$ in a set of maps.
In this method, we chose the
minimal amplitude of $\Delta T_p$ in the pixel as the CMB signal,
i.e. $L_{\min}(\{\Delta T^{(i)}_p \}) \rightarrow \Delta T^{\min}_p$,
where $|\Delta  T^{\min}_p|=\min\{|\Delta T^{(i)}_p|\}$.
To estimate residues from point sources,
we find the maximal amplitude of the signal in each pixel
$L_{\max}(\{\Delta T^{(i)}_p\}) \rightarrow \Delta T^{\max}_p$,
where $|\Delta  T^{\max}_p|=\max\{|\Delta T^{(i)}_p|\}$,
for all the pre-CMB maps produced by the method Eq.(\ref{eq9}).
For true CMB, the difference $\Delta T^{\max}_p-\Delta T^{\min}_p$
is equal to zero and we consider
$\Delta T^{\max}_p-\Delta T^{\min}_p$
as a measure of deviation from the CMB map.

The reason for such a filter is quite obvious. The signal in
each pixel is a superposition of the CMB signal and (small) 
residues of the foregrounds. If correlations between the CMB
signal and foregrounds are minimal then we expect that
all the deviations of the $\Delta T^{(i)}_p$ in the pixel are caused
by the residues. In our case, two approximations of the CMB map
restored by Ka--V and Q--V channels (Naselsky et al. 2003)
have been used
and $\Delta T^{\min}$ has been taken as the CMB map.

The result of the PCM application to the {\it WMAP}
data at the range $\ell\le100$ is shown in Fig.\,1 for the CMB signal.
In this Figure,
the differences between the ILC and PCM maps
(top right), TOH FCM and PCM (bottom left) and
between TOH Wiener and PCM maps (bottom right) are also shown.
As is shown in Fig.\,1,
the main difference is connected with sources in the Galactic
plane and the harmonic $a_{51}$ which is probably related
to the galactic emission. The difference shows foreground components 
being absent in the PCM map.

For the simplest estimator of the power spectrum,
$C(\ell)=(2\ell+1)^{-1}\sum_m |\alm|^2$,
we show in Fig.\,2 the power spectrum for the ILC, FCM, PCM and TOH
Wiener filtered maps. As one can see from Fig.2
the PCM map reproduces well the spectrum for the TOH Wiener
filtered map.

\begin{figure}
\centering
\epsfxsize=8.cm
\epsfbox{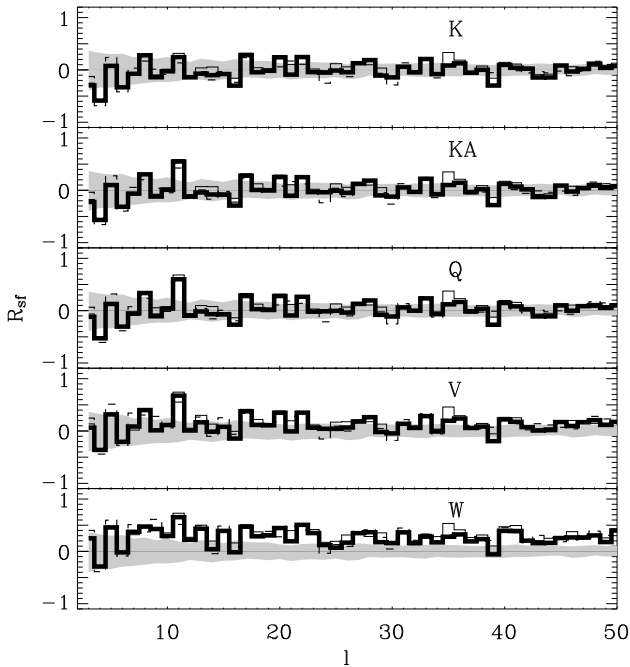}
\vspace{1.cm}
\caption{
The circular correlation between the
ILC (thick solid lines), FCM (thin solid lines) and PCM (dashed lines)
cleaned signals and derived foregrounds
for K--W channels.
} 
\label{fc} 
\end{figure}

\section{Correlations of the cleaned signals and foregrounds 
for the ILC, FCM and PCM maps}

Three cleaned maps (ILC, FCM and PCM) are produced from different
methods of the foreground component separation. For the
perfect separation of the CMB signal and foregrounds we can 
expect small random correlations between their phases. However, 
these methods provide the approximate separation only.
Therefore, these remaining cross--correlations
between phases of the cleaned signal, $\phi_s$, and 
the foregrounds, $\psi_f$, can be used to characterize the degree
of separation achieved.

To do this, here we will use the
simplest phase--phase circular (Fisher, 1993) correlation 
coefficients.  
We will use also the linear correlation coefficients 
and more refined ``friend--of--friend'' statistics discussed, 
for example, by Roeder (1992) neglecting circularity of the phases.
Indeed, equivalence of the phases $\phi_s\,\&\,
\phi_s\pm 2\pi$ and $\psi_f\,\&\,\psi_f\pm 2\pi$, changes 
their correlation functions and cumulants and
the results obtained with the ``friend--of--friend'' statistics.
To suppress this effect we compared our result
with 200 random realizations of Poissonian process
prepared in the same manner (see Sec.\,3.3).

\begin{figure}
\centering
\epsfxsize=8.cm
\epsfbox{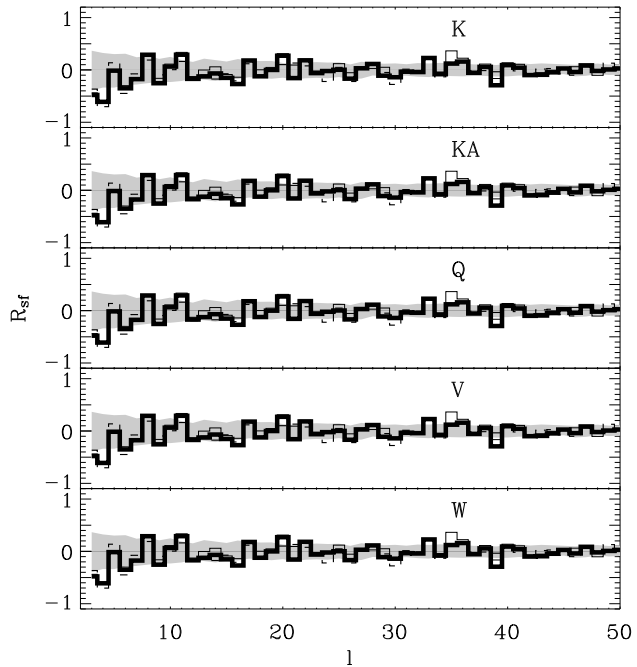}
\vspace{1.cm}
\caption{
The circular correlation between the
ILC (thick solid lines), FCM (thin solid lines) and PCM (dashed lines)
cleaned signals and free--free emission
for K--W channels.
} 
\label{fcs}
\end{figure}

For five frequency channels K--W, the maps are taken from the
{\it WMAP}
web site\footnote{\tt http://lambda.gsfc.nasa.gov/product/map/m\_products.cfm},
and all the phases
are obtained by the spherical harmonics decomposition
using the HEALPix (G\'orski et al. 1999) and
GLESP (Doroshkevich et al. 2003a)
codes. We consider separately two ranges of multipoles, 
$2\leq\ell\leq 50$ and $51\le\ell\le 100$.
For the ILC, FCM and PCM CMB maps, the derived foregrounds
are determined as differences between the signal ($S$) and
cleaned map ($C$) $F=S-C$ for each ($\ell,m$).
In addition, for
the ILC cleaned map, {\it WMAP} own foregrounds given in the same
web--site were also used. 
Examples of difference of $S$, $ILC$ and WMAP own foregrounds for
V and W WMAP bands are shown on Fig.\,3.

\subsection{Circular cross-correlation of the phases.}

 Following Fisher(1993) we define the statistics
\[
M_{sp}={1\over \ell_{max}}\sum_m^{\ell_{max}}\exp[ip(\phi_m-\langle\phi_m
\rangle)]\,,
\] 
\[
M_{fp}={1\over \ell_{max}}\sum_m^{\ell_{max}}\exp[ip(\psi_m-\langle\psi_m
\rangle)]\,,
\]  
\[
\langle\phi_m\rangle=\tan^{-1}\left(\sum_m^{\ell_{max}}\sin\phi_m/
\sum_m^{\ell_{max}}\cos\phi_m\right)\,,
\]
\[
\langle\psi_m\rangle=\tan^{-1}\left(\sum_m^{\ell_{max}}\sin\psi_m/
\sum_m^{\ell_{max}}\cos\psi_m\right)\,,
\]

\begin{equation}
R_{sf}(\ell)={1\over \ell}\sum_{m=1}^\ell \cos(\phi_m-\psi_m)\,,
\label{circ}
\end{equation}
\[
\langle R_{sf}\rangle={1\over \ell_{max}-\ell_{min}+1}\sum_{\ell=\ell_{min}}^{\ell_{max}} R_{sf}(\ell)
\]
 where $M_{sp}$ and $M_{fp}$ are the $p$-th trigonometric moments of
the samples, $\langle\phi\rangle$ and $\langle\psi\rangle$ are 
corresponding mean directions, $R_{sf}(\ell)$ is the circular
cross-correlation coefficient in each mode $\ell$ and $r_{sf}$ is
the mean circular cross-correlation coefficient for all phases.
For $m=0$ and for all $\ell$ phases $\phi(\ell,0)=\psi(\ell,0)=0$
and here we neglect them. 

\begin{figure}
\centering
\epsfxsize=8.cm
\epsfbox{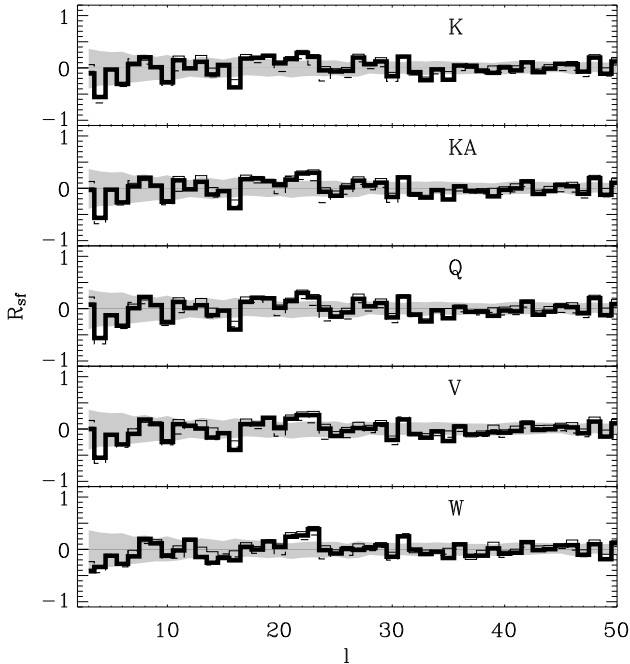}
\vspace{1.cm}
\caption{
The circular correlation between the
ILC (thick solid lines), FCM (thin solid lines) and PCM (dashed lines)
cleaned signals and synchrotron emission
for K--W channels.
} 
\label{fcf}
\end{figure}

\begin{figure}
\centering
\epsfxsize=8.cm
\epsfbox{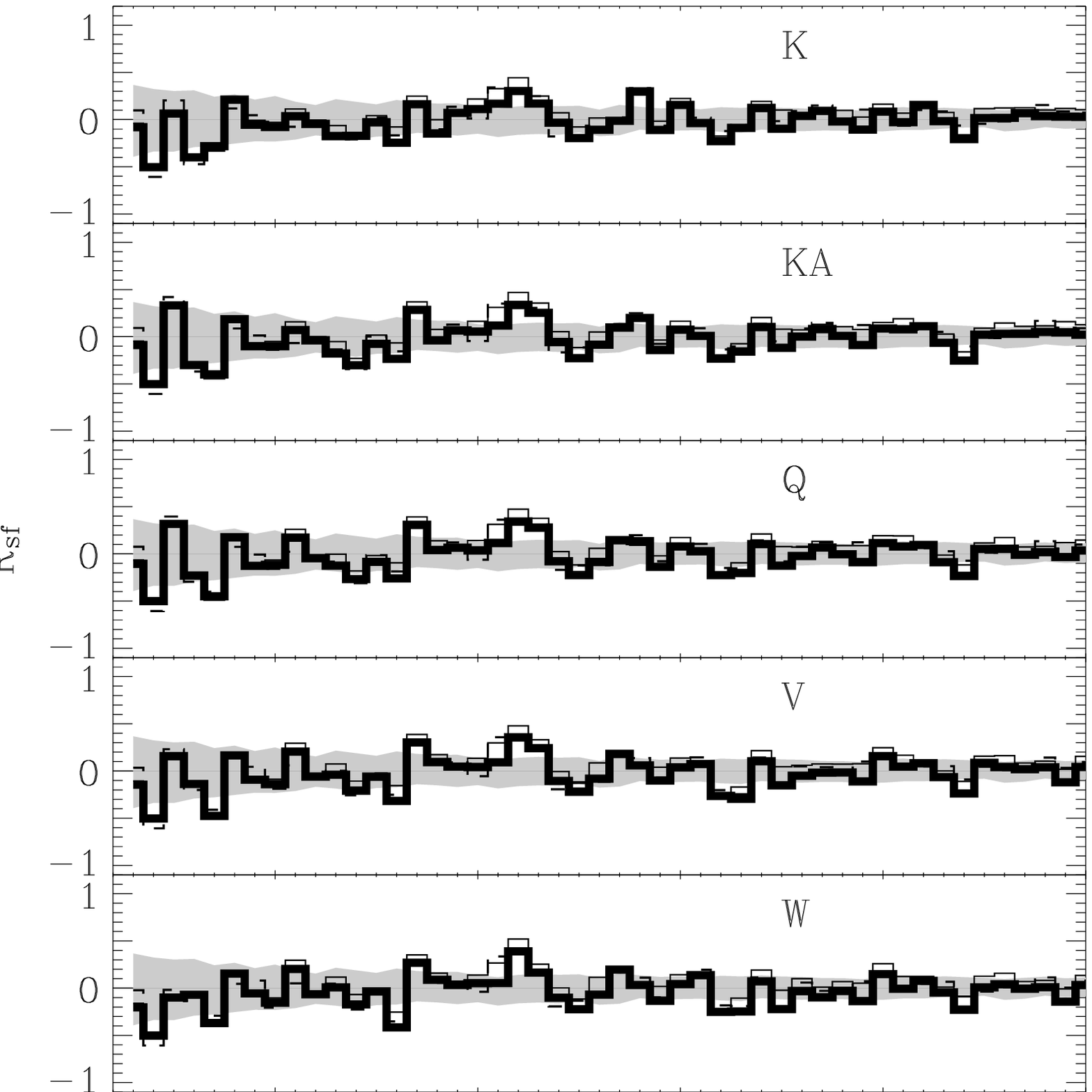}
\vspace{1.cm}
\caption{
The circular correlation between the
ILC (thick solid lines), FCM (thin solid lines) and PCM (dashed lines)
cleaned signals and dust emission
for K--W channels.
} 
\label{fcd}
\end{figure}

\begin{table}
\begin{tabular}{l rrr rrr} 
\hline
&K~~~~&KA~~&Q~~~~&V~~~~&W~~~~\cr
\hline
$ILC^{(o)}$&-0.026&-0.031&-0.030&-0.033&-0.033\cr
$ILC^{(d)}$&-0.017&0.018&0.022&0.112&0.262\cr
$FCM$&0.031&0.051&0.071&0.157&0.320\cr
$PCM$&-0.019&0.007&0.032&0.136&0.288\cr
\hline
\end{tabular}
\caption{Circular cross-correlation coefficients, $\langle R_{sf}\rangle$,
between phases of the ILC signals and own and derived foregrounds, 
and PCM and FCM signals with their derived foregrounds.
}
\label{Tab11}
\end{table}

For the K--W bands the circular coefficients, $R_{sf}(\ell)$, are
plotted in Fig.\,4 for the ILC cleaned signal and its own and
derived foregrounds and for PCM signal and derived foregrounds.
For the first three channels, these coefficients are quite moderate
and do not exceed the random scatter ($1\sigma$) obtained from 
200 random realizations. Note that, for all the bands, the shape 
of the functions $R_{sf}(\ell)$ are quite similar to each other what
reflects strong correlation of phases in all the foregrounds 
(Naselsky et al. 2003). As one can see from Fig.\,4, for the
V\,\&\,W channels the cross--correlations of both the PCM and ILC phases
with the derived foregrounds seems to be quite significant. 

The same tendency is seen from the estimations of the mean 
coefficients, $\langle R_{sf}\rangle$,
for $2\leq \ell\leq 50$ listed in Table\,1
for all three cleaned signal and the own WMAP and three derived 
foregrounds. For first three channels these coefficients are 
small but they become significant for channels V and W and
derived foregrounds.

Correlations of phases of different foregrounds with the ILC, FCM 
and PCM cleaned signals are plotted in Figs. \,5--7.
For the free--free foregrounds, correlations are quite moderate and
they exceed the random scatter only for the 35 -- 36 harmonics.
However, for the
synchrotron and dust foregrounds we see the significant (95\%)
correlations for the 21 -- 23 harmonics
for all cleaned
signal maps.
In all the cleaned maps this range corresponds to minima of the power
(see Fig.\.2).

\begin{table*}
\begin{minipage}{150mm}
\begin{tabular}{lcc ccc ccc c} 
\hline
channel&W-ILC&W-FCM&W-PCM&ILC-50&ILC-100&FCM-50&FCM-100&PCM-50&PCM-100 \cr
\hline
K &0.022&0.019&0.026&0.032&0.047&0.035&0.022&0.042&0.056\cr
KA&0.028&0.032&0.041&0.034&0.046&0.030&0.027&0.059&0.048\cr
Q &0.028&0.034&0.045&0.054&0.069&0.048&0.040&0.065&0.074\cr
V &0.023&0.028&0.051&0.073&0.116&0.081&0.088&0.113&0.121\cr
W &0.033&0.033&0.057&0.143&0.191&0.142&0.167&0.194&0.209\cr
\hline
\end{tabular}
\caption{Linear correlation coefficients between the phases of
the foregrounds and cleaned signal for all the K--W bands.
The first three columns represent $r_{sf}$ for {\it WMAP} 
own foregrounds and the ILC, FCM and PCM cleaned signals. 
In other columns the cross-correlation coefficient for ILC, 
FCM and PCM signals with their derived foregrounds are 
listed for two ranges of multipoles, $\ell\le50$ and 
$51\le\ell\le 100$.
}
\end{minipage}
\label{Tab12}
\end{table*}

\subsection{Linear cross-correlation of the phases}

For $\ell_{min}\leq\ell\leq\ell_{max}$, the linear correlation
coefficient between phases of the cleaned signal, 
$\phi_s$, and foregrounds $\psi_f$, is defined as follows:
\begin{equation}
r_{sf}(\ell_{min},\ell_{max})={\langle\phi_s\psi_f\rangle-
\langle\phi_s\rangle\langle\psi_f\rangle\over 
\sigma_\phi\sigma_\psi}\,,
\label{rsf}
\end{equation}
\[
\langle\phi_s\rangle={1\over N_\ell}\sum_{\ell=\ell_{min}}^{\ell_{max}}
\sum_{m=1}^{\ell}\phi_s(\ell,m)\approx\pi\,,
\]
\[
\sigma_\phi^2={1\over N_\ell}\sum_{\ell=\ell_{min}}^{\ell_{max}}
\sum_{m=1}^{\ell}\phi_s^2(\ell,m)\approx \pi^2/3,
\]
\[
\langle\phi_s\psi_f\rangle={1\over N_\ell}\sum_{\ell=\ell_{min}}^{\ell_{max}}
\sum_{m=1}^{\ell}\phi_s(\ell,m)\psi_f(\ell,m)\,,
\]
\[ 
N_\ell=(\ell_{max}-\ell_{min}+1)(\ell_{max}+\ell_{min})/2\,.
\]
For the foreground phases, $\psi_f$, the mean value, 
$\langle\psi_f\rangle$, and the variance, $\sigma_\psi$, 
are defined with similar relations. Here $N_\ell$ is the 
number of phases in the sample under consideration. For 
$m=0$ and for all $\ell$, phases $\phi_s(\ell,0)=
\psi_f(\ell,0)=0$ and here we neglect them. 

The linear correlation coefficients for $2\leq\ell\leq 50$
and $51\leq\ell\leq 100$ are listed in Table\,2 for the ILC,
FCM and PCM cleaned signals and five frequency channels. 
First three columns characterize the correlation with the  
ILC own foregrounds. Six other columns represent these 
coefficients for the same signals and derived foregrounds.  
For six samples the correlations between the signal and 
derived foregrounds, $r_{sf}(\ell_{min},\ell_{max})$, 
increase with the channel frequency and are comparable 
for a given frequency. The random scatter of the coefficient 
determined by averaging of 200 random realizations is
$r_{sf}(\ell_{min},\ell_{max})\sim 0.06$ at $68\%$ CL. 
As is seen from Table\,1, for high frequency channels 
V and W the measured $r_{sf}(\ell_{min},\ell_{max})$ 
exceeds this value. This fact indicates the noticeable 
correlation between the cleaned signal and derived 
foregrounds in channels V and W and the limited precision 
of separation methods. 

For all the channels and {\it WMAP} own foregrounds, 
the coefficients $r_{sf}(\ell_{min},\ell_{max})$
are less than the random value that indicates high 
efficiency of corrections of own foregrounds. However, 
this result depends upon the foregrounds rather than 
the cleaned signal. For the FCM and PCM cleaned signal 
and the same foregrounds, the coefficients, $r_{sf}
(\ell_{min},\ell_{max})$ are also listed in the same 
Table. For all samples these coefficients $r_{sf}
(\ell_{min},\ell_{max})$ do not exceed the random value. 
Therefore, the small correlations between the own 
foregrounds of ILC and the cleaned signals is 
determined by the properties of foregrounds.

By definition, this coefficient is more sensitive to 
$\ell_{max}$ because the majority of phases used comes
from $\ell\sim\ell_{max}$. However, it decreases when we
correlate phases with different $\ell$. 

\subsection{Linear correlation coefficient of the 
phases per each multipole $\ell$}

Linear correlation coefficient of phases of cleaned signal and 
foregrounds, $r_{sf}(\ell)$, can also be found for each $\ell$ 
in the same manner as it was been done for the circular cross--
correlation coefficient in Sec. 3.1. In the case, the coefficient 
is defined by the same relations (\ref{rsf}) neglecting the 
summation over $\ell$ and for $N_\ell=\ell$. It allows one to
characterize properties of each mode separately and to determine 
harmonics with maximal $r_{sf}(\ell)$. However, the small sample 
statistics of phases at moderate $\ell$ increases significantly 
its random scatter. Note also that, as is seen from the definition, 
this coefficient characterizes the sample used. Thus, in contrast 
with the circular cross--correlation coefficient, the direct 
averaging $r_{sf}(\ell)$ over $\ell$ does not reproduce results
obtained in previous Section. 

\begin{figure}
\centering
\epsfxsize=7.cm
\epsfbox{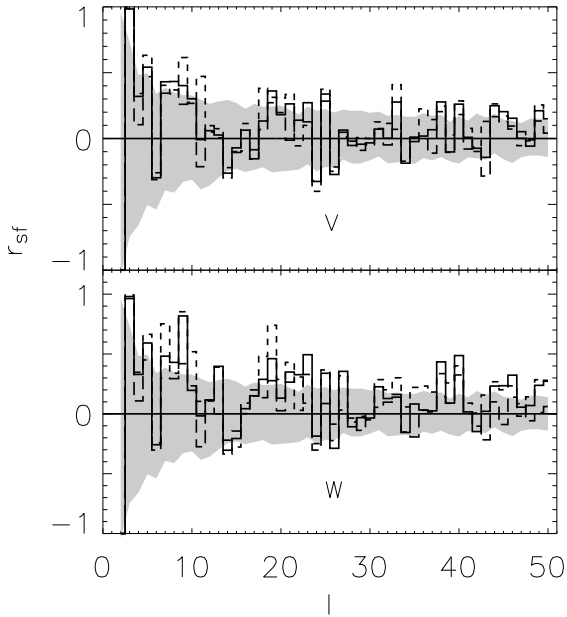}
\vspace{1.cm}
\caption{Linear correlation coefficient, $r_{sf}$, between the CMB
and foregrounds vs. the harmonics number $\ell$, in V and W channels
for  ILC (solid line), ILC and its own foregrounds (dash line, which
overlapped practically the solid line), and PCM filtered map (long dashed line).
} 
\label{f22}
\end{figure}

The coefficients $r_{sf}(\ell)$ are plotted in Fig. \ref{f22} for
channels V and W for two foregrounds of the ILC cleaned
signal and for the PCM cleaned signal. For each $\ell$ the
error bars are found for 200 random realizations and are
well consistent with the expected behavior $\approx \ell^{-1/2}$.

For channels K--V and for all three samples,
the correlation coefficients do not exceed the random scatter
but the disposition of their maxima and minima are very 
similar. For the channel W
we see significant differences between functions $r_{sf}(\ell)$
for different samples. More pronounced correlations are seen for
$4\leq\ell\leq 10$, $\ell=13 - 15$, $\ell=18 - 27$, $\ell=32 - 36$
and $\ell=45 - 50$
As is seen from the comparison with
Fig.\,\ref{f2}, some of these harmonics correspond to local minima
in $C_{\ell}$. This fact indicates that, perhaps, these
peculiarities are caused by the same sources as
the Galactic dust emission.

\subsection{``Friend--of--Friend'' statistics}

To compare the foregrounds and the foreground-cleaned signal
with random Poissonian process we use a more refined
technique, the ``friend--of--friend'' approach
for the 2D phase diagramme. In these diagrammes each point
has coordinates $x_i=\phi_s(\ell,m), y_i=\psi_f(\ell,m)$
where $\phi_s(\ell,m)$ and $\psi_f(\ell,m)$ are
the phases of the cleaned signal and the foregrounds, respectively.

For each diagramme, we find the fraction of clusters, $f_M$, with a
given richness, $M$, in wide range of linking lengths, $r_{lnk}$, 
which determines the maximal point separation within a cluster. 
Here we use the normalized linking length, 
\begin{equation}
l_{lnk}=\pi r_{lnk}^2\langle n\rangle,\quad \langle n\rangle=
N_{phs}/4\pi^2\,,
\label{lnk}
\end{equation}
where $r_{lnk}$ is the actual distance between two points,
and $N_{phs}$ is the number of points used.
 As was
discussed in White (1979) and Borgani (1996), the fractions 
$f_M$ depend on the correlation functions (or cumulants) of 
all orders that provides high sensitivity to deviations of 
the sample from the Poissonian--like one. This sensitivity is
weak for $M=1\,\&\,2$ and fast increases for larger $M$. 
But the random scatter of $f_M$ increases for larger $M$ as well,
and the analysis becomes ineffective for $M\geq 7$.

For the truly Poissonian sample of points, the correlations
functions of all orders are equal to zero and these fractions 
can be approximately described as follows:
\[
f_M(l_{lnk})\approx [1-\exp(-l_{lnk})]^M\exp(-l_{lnk})\,,
\]
where the first term describes the probability to find $M$ separation 
between points $\leq l_{lnk}$ while the last term gives probability
to find larger separation for the $M+1$ point. The scatter of $f_M$ 
is determined by averaging of $f_M$ over 200 random realizations of
diagrammes analyzed in the same manner.

To find clusters with a given richness we use the Minimal Spanning 
Tree technique which, firstly, allows one to connect all points within
a unique tree.
Further on, rejection of edges of a tree larger than a
chosen linking length transforms the tree to the system of clusters of
all richness for a given linking length.
The Minimal Spanning Tree technique was proposed
in Barrow et al. (1985) and van de Weygaert (1991) and was applied 
for analysis of rich galaxy catalogues (see, e.g., Doroshkevich et al. 
2001, 2003b).

Firstly, we are testing the hypothesis that
the distribution of phases are Poissonian-like and
correlations are negligible for both foregrounds and cleaned signal.
Main results of our analysis are
presented in Figs.\,9,10 where for all five frequency channels
variations of fractions $f_M(l_{lnk}), 1\leq M\leq 7$ are plotted 
versus the linking length, $l_{lnk}$. As is expected,
the functions  $f_M(l_{lnk})$ are quite similar to
Poissonian ones
for $M=1$ and $M=2$,
and even for $M=3$ the deviations are moderate.
However, for $M\geq 4$ the deviations from the Poissonian samples
become significant, especially for V and W channels. This result
illustrates the expected non-Gaussian character of the cleaned 
signal and its correlations with the foregrounds. 

This result is expectable because of non-Gaussian character 
of the foregrounds signals. However, comparing the phase 
diagramme for the cleaned signal and foregrounds with 100
diagrammes prepared for the foreground and random phases
we can quantify their divergences using $\chi^2$ 
statistics. These results are shown 
in Table~3. If for $M=1$ and $M=2$ we see small
$\chi^2\leq N$  where $N$ is the number of linking lengths 
used then for $M\geq 3$, we have $\chi^2\geq N$. Larger
ratio $\chi^2/N$ is found again for higher frequency 
channels V and W. This fact indicates that all
the cleaned maps contain some residues from the dust
emission. 

\begin{figure*}
\begin{minipage}{160mm}
\centering
\vbox{
\epsfxsize=15.cm
\epsfbox{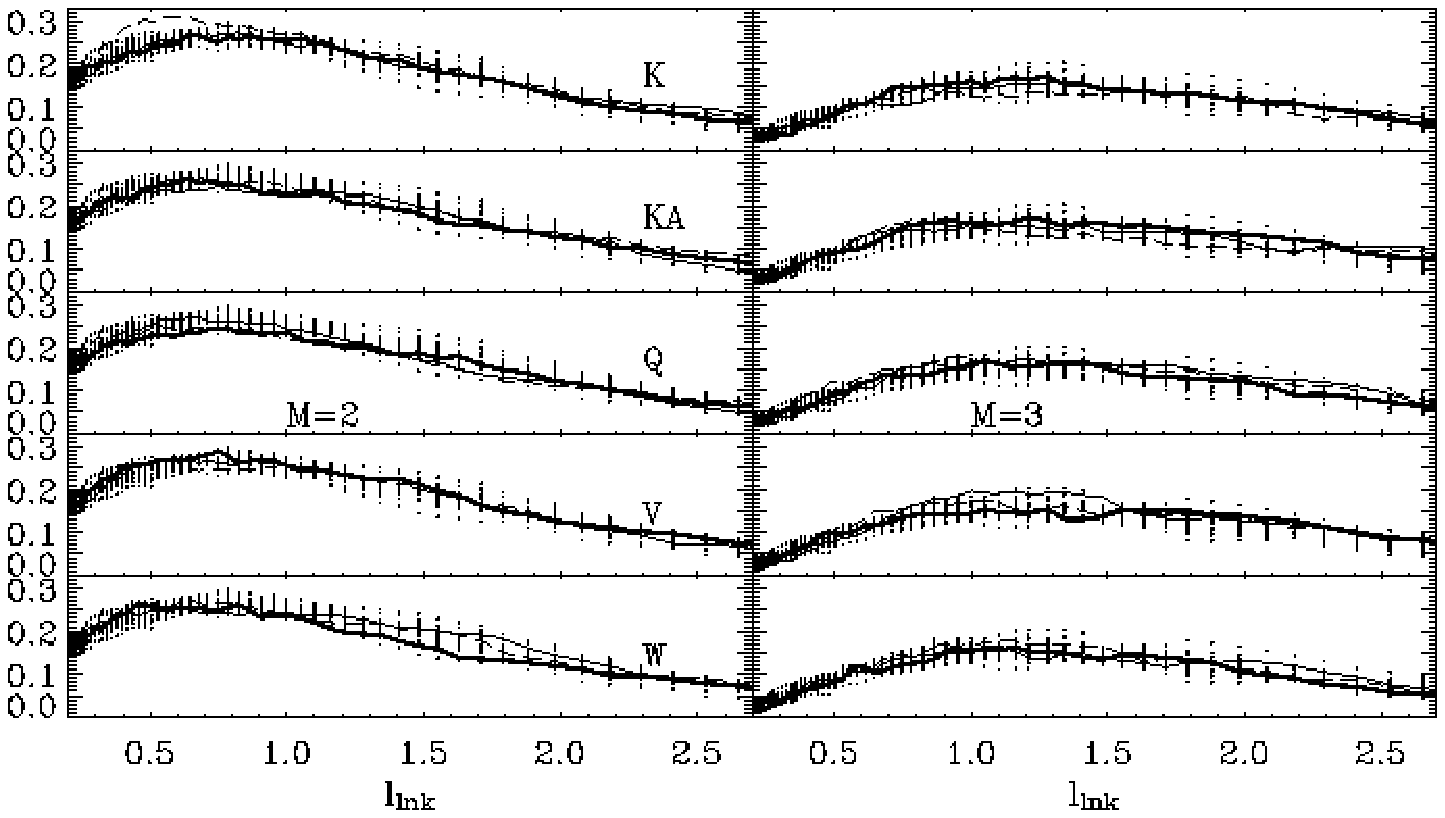}
\vspace{0.1cm} }
\vbox{
\epsfxsize=15.cm
\epsfbox{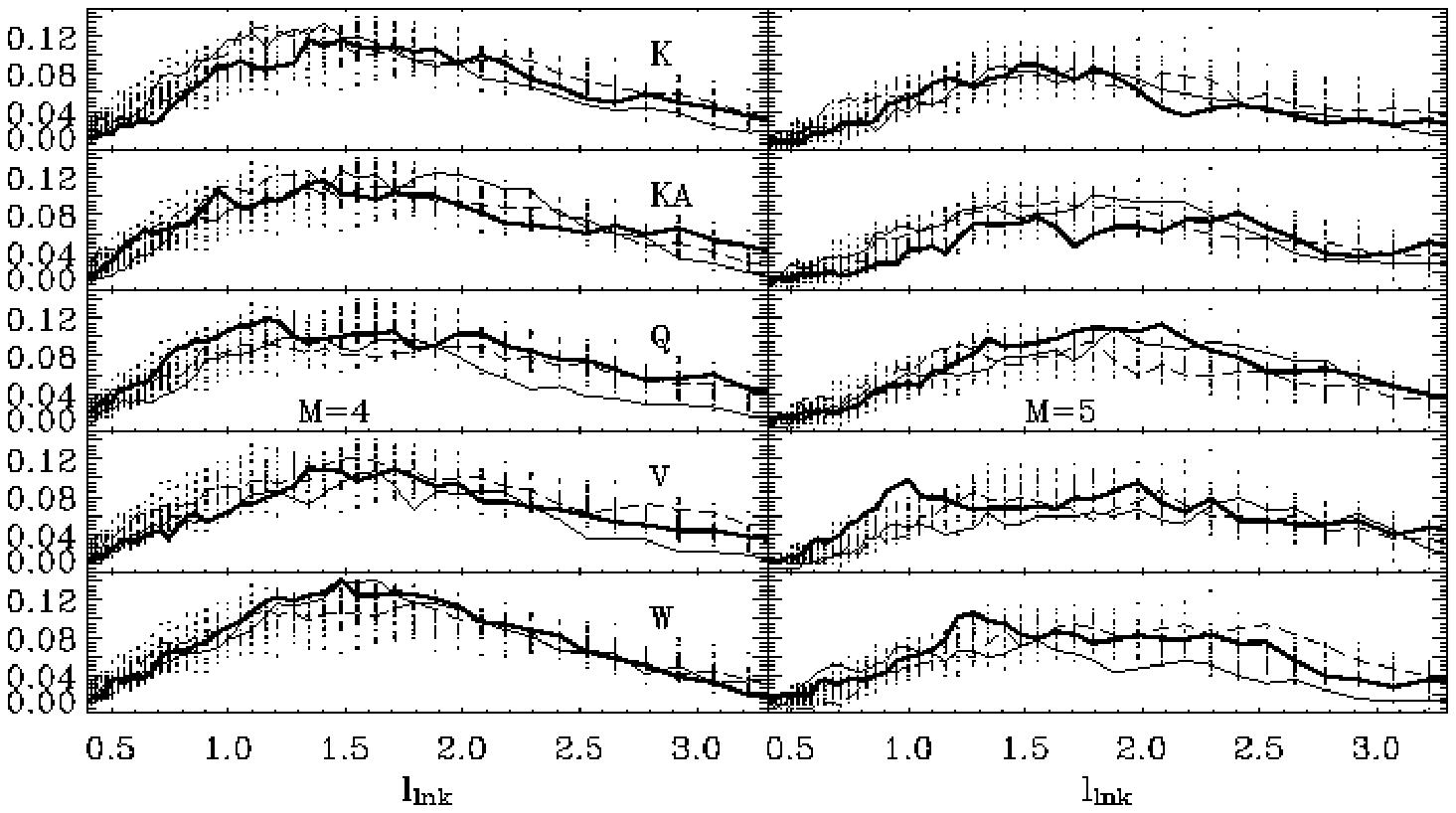}
\vspace{0.1cm} }
\vbox{
\epsfxsize=15.cm
\epsfbox{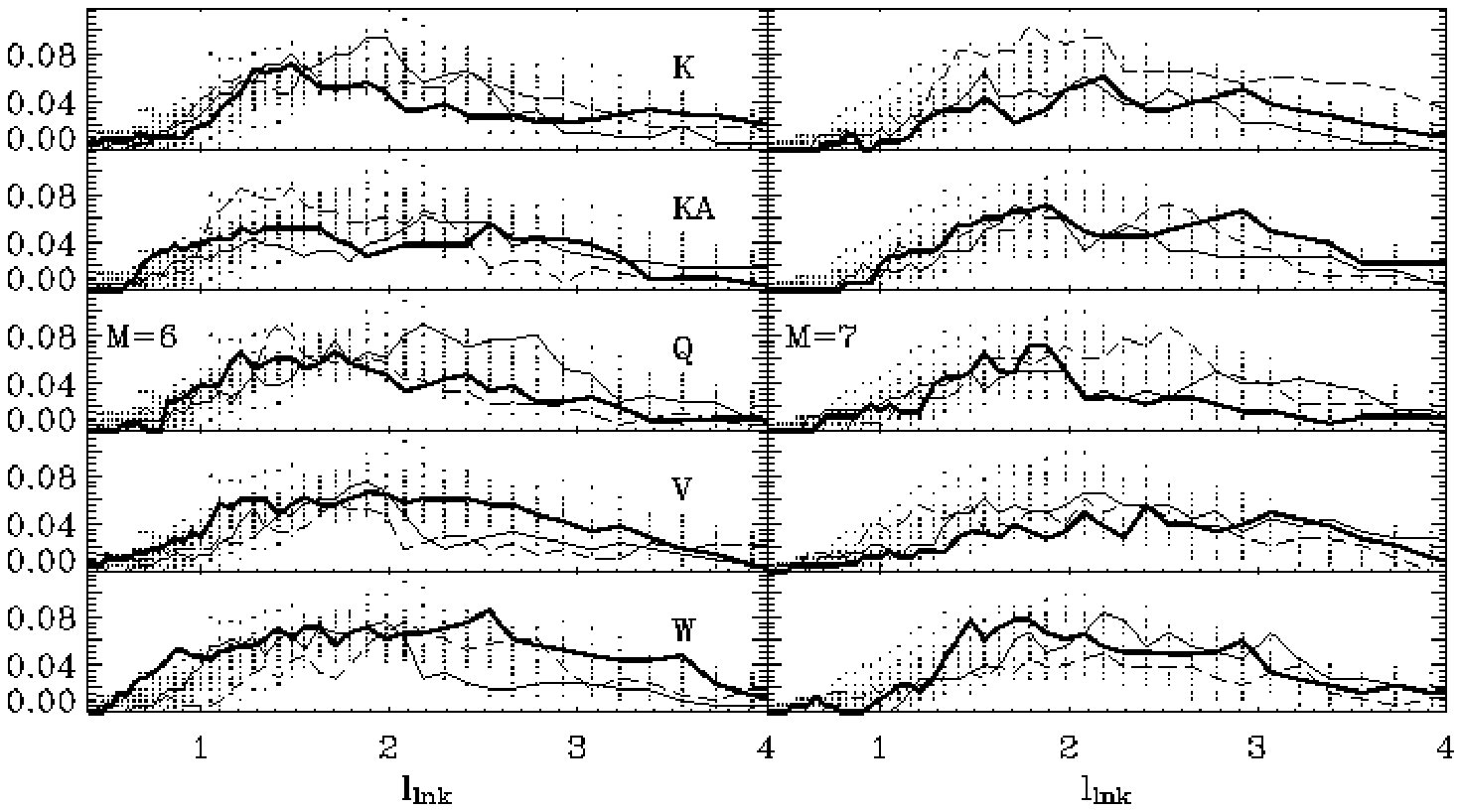}
\vspace{0.1cm} }
\caption{Fraction of clusters with richness
$M=2-7$ vs. dimensionless linking
length, $l_{lnk}$ for the PCM reconstructed CMB (thick solid line), 
ILC (dash line), TOH FCM (thin solid line). The dots represent a set of
random samples. The range of multipoles for the phases is 
$2\le\ell\le 50$.
} 
\label{ff2}
\end{minipage}
\end{figure*}

\begin{figure*}
\begin{minipage}{160mm}
\centering
\vbox{
\epsfxsize=12.cm
\vspace{1.cm}
\epsfbox{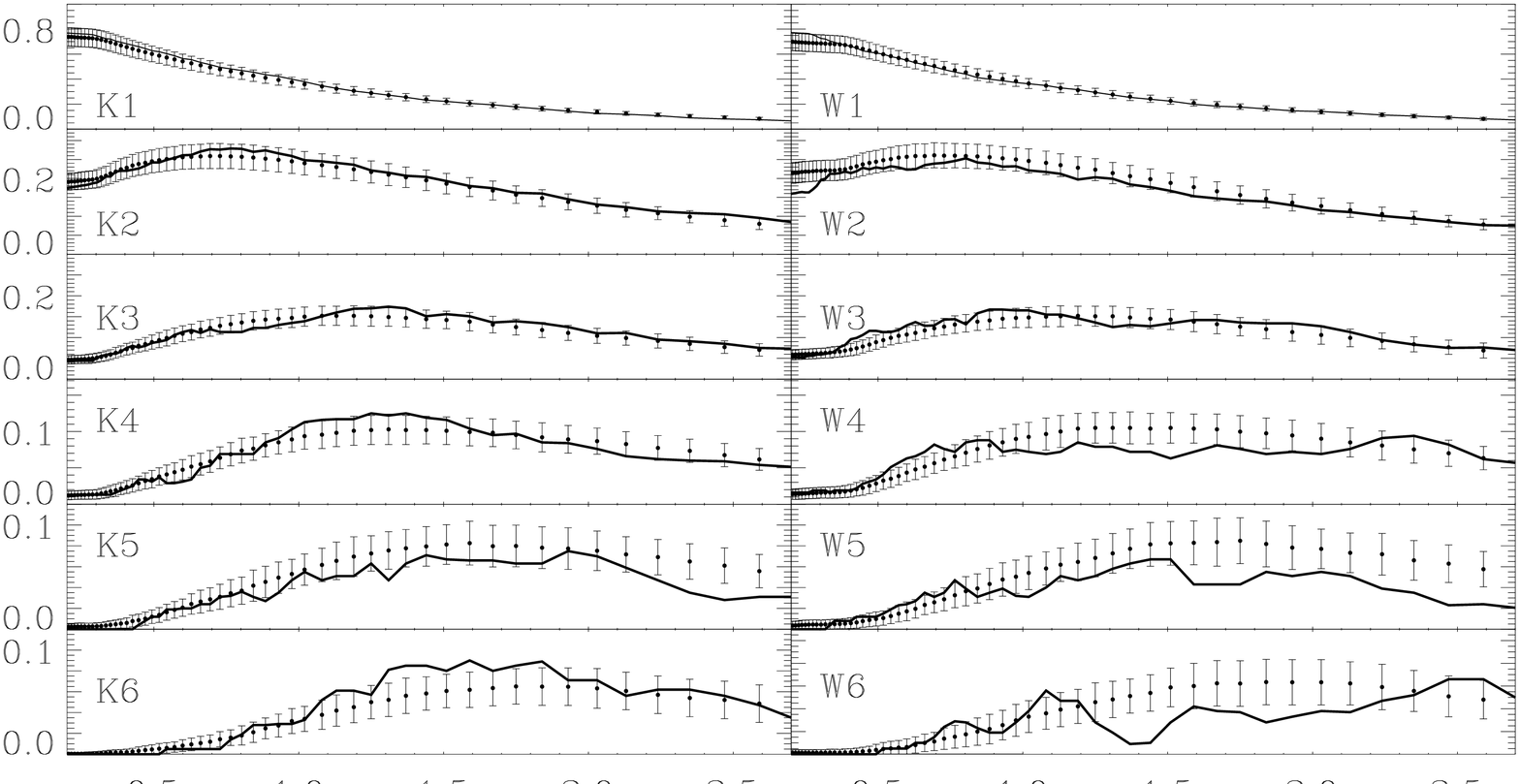}
\vspace{1.cm}}
\vbox{
\epsfxsize=12.cm
\epsfbox{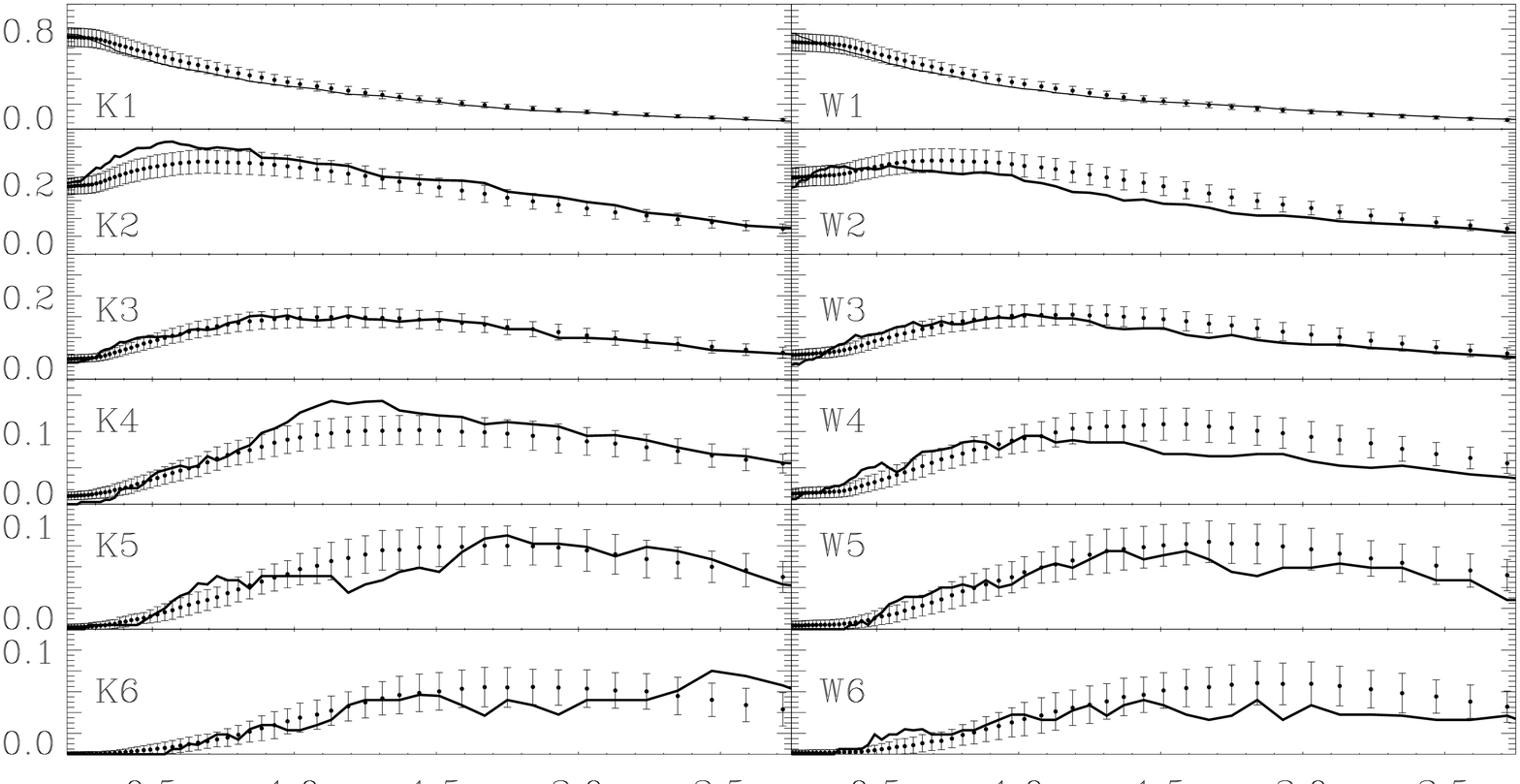}
\vspace{1.cm}}
\vbox{
\epsfxsize=12.cm
\epsfbox{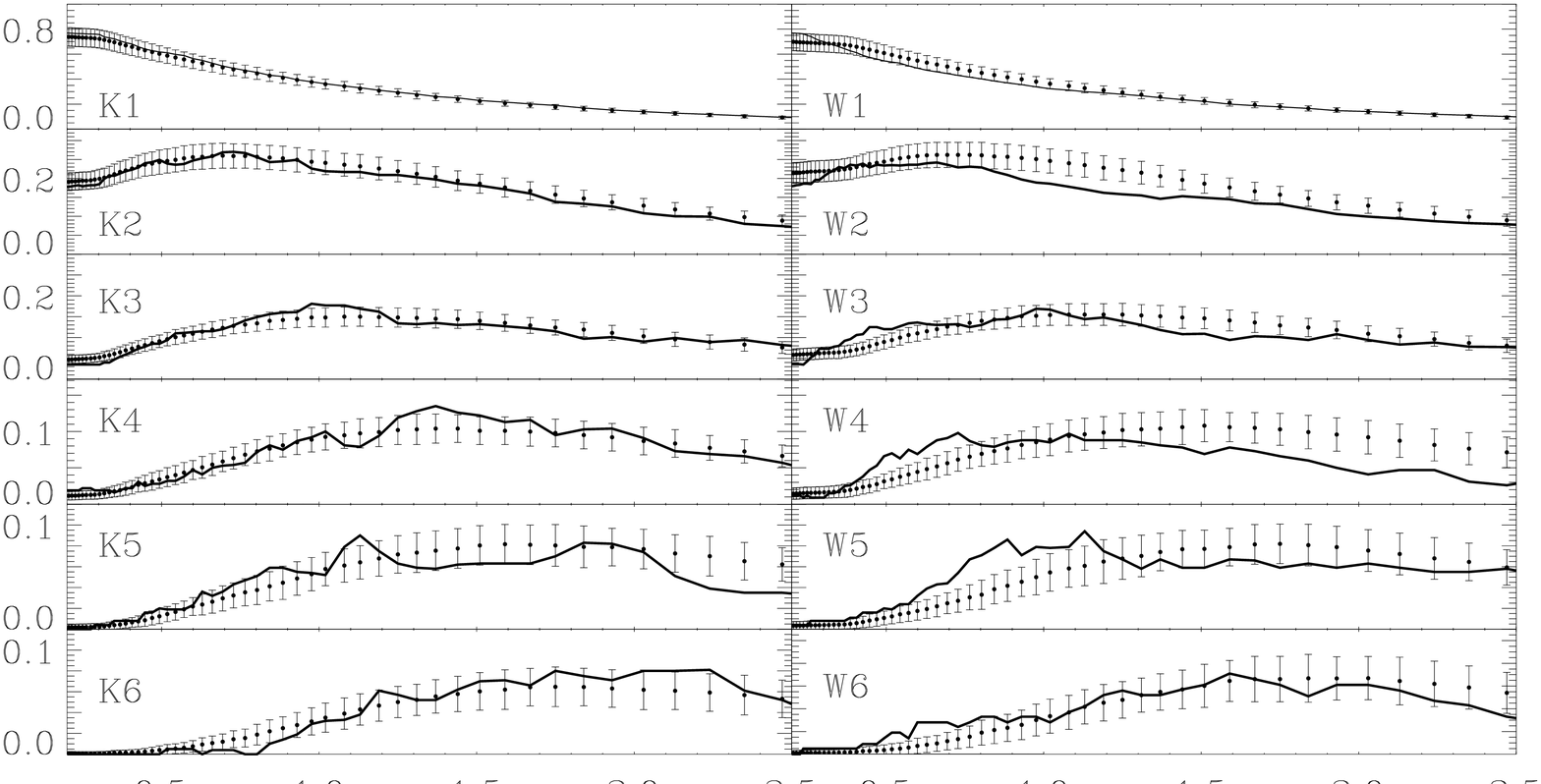}
\vspace{0.7cm}}
\caption{Fraction of clusters with richness 
$M=2-6$ for K and W channels vs. dimensionless 
linking length, $l_{lnk}$ for the ILC (top panels), TOH FCM (middle panels)
and PCM (bottom panels) reconstructed CMB. The range of multipoles
used is $2\le\ell\le 50$. The error bars corresponds
to $1\sigma$ level.
} 
\label{ff67}
\end{minipage}
\end{figure*}

\begin{table*}
\begin{minipage}{160mm}
\caption{
Comparison of
degrees of clusterization between the foreground and random 
phases (200 realization) and the foreground and CMB phases
for five channels of the ILC, FCM and PCM  samples.
The $\chi^2$ and
the number of linking lengths used are presented
for clusters with richness between 1 and 7.
}
\begin{tabular}{lrr rrr rrr rr rrr rr} 
\hline
channel&$f_1$&$f_2$&$f_3$&$f_4$&$f_5$&$f_6$&$f_7$&\qquad&$f_1$&$f_2$&$f_3$&$f_4$&$f_5$&$f_6$&$f_7$\cr
\hline
\multicolumn{1}{c} {}         &
\multicolumn{7}{c} {ILC--50}  &
\multicolumn{7}{c} {\hspace*{1cm}ILC--100} \cr
   K&    72&   79&   94&   59&   46&   40&   36&\qquad\qquad& 75&   60&   74&   56&   45&   40&   34\cr 
$\chi^2$& 5&   11&   24&   36&   52&   34&   29&\qquad\qquad&  9&   13&   78&   28&   28&   24&    9\cr 
  KA&    71&   78&   93&   56&   45&   40&   34&\qquad\qquad& 75&   60&   73&   56&   45&   38&   34\cr 
$\chi^2$&18&   20&   68&   53&   33&   38&   33&\qquad\qquad&  3&    7&   37&   58&   21&   16&   47\cr 
   Q&    72&   79&   94&   57&   46&   39&   34&\qquad\qquad& 75&   60&   74&   56&   45&   38&   34\cr 
$\chi^2$& 2&   31&  160&   42&   21&   24&   39&\qquad\qquad&  2&    4&   42&   69&   43&   84&   35\cr 
   V&    72&   79&   94&   57&   46&   39&   34&\qquad\qquad& 75&   60&   74&   56&   45&   39&   34\cr 
$\chi^2$& 2&   29&   29&   50&   35&   27&   11&\qquad\qquad&  2&    8&   78&   34&  120&   38&   36\cr 
   W&    71&   78&   92&   58&   46&   39&   34&\qquad\qquad& 75&   59&   74&   57&   46&   40&   34\cr 
$\chi^2$& 2&   69&  145&   59&   78&   63&   51&\qquad\qquad& 11&   49&   58&   58&   99&   60&   47\cr 
\hline
\multicolumn{1}{c} {}         &
\multicolumn{7}{c} {FCM--50}  &
\multicolumn{7}{c} {\hspace*{1cm}FCM--100} \cr
   K&      71&   78&   93&   57&   47&   40&   35&\qquad\qquad& 75&   60&   74&   56&   45&   40&  34\cr 
$\chi^2$&  22&  134&   47&   52&   42&   31&  122&\qquad\qquad&  2&    9&   28&   40&   15&   26&  14\cr 
  KA    &  72&   79&   93&   56&   47&   40&   35&\qquad\qquad& 75&   60&   74&   56&   46&   39&  34\cr 
$\chi^2$&  10&   90&  150&   57&   55&   21&   26&\qquad\qquad&  2&    3&   41&   60&    9&   34&  25\cr 
   Q&      71&   78&   93&   56&   46&   40&   35&\qquad\qquad& 75&   60&   74&   56&   45&   40&  34\cr 
$\chi^2$&   9&    8&  137&   17&   85&   12&   18&\qquad\qquad&  4&    6&   23&   26&   18&   26&  18\cr 
   V&      71&   78&   93&   56&   46&   40&   35&\qquad\qquad& 75&   60&   74&   56&   45&   39&  34\cr 
$\chi^2$&   5&   40&  115&   75&  117&  111&   37&\qquad\qquad&  1&   19&   32&   49&   22&   18&  55\cr 
   W&      72&   79&   94&   57&   46&   39&   34&\qquad\qquad& 75&   60&   74&   56&   45&   38&  34\cr 
$\chi^2$&  24&   84&   67&   84&   36&   35&   52&\qquad\qquad& 10&    5&   27&   63&   32&   86&  36\cr 
\hline
\multicolumn{1}{c} {}         &
\multicolumn{7}{c} {PCM--50}  &
\multicolumn{7}{c} {\hspace*{1cm}PCM--100} \cr
   K&      72&   79&   94&   58&   47&   39&   36&\qquad\qquad&  75&   60&   74&   56&   45&   40&   34\cr 
$\chi^2$&   2&    6&   37&   33&   70&   28&   13&\qquad\qquad&   9&   13&   78&   28&   28&   24&    9\cr 
  KA&      72&   79&   94&   58&   47&   40&   35&\qquad\qquad&  75&   60&   73&   56&   45&   38&   34\cr 
$\chi^2$&   2&   13&   32&   83&   95&  125&   16&\qquad\qquad&   3&    7&   37&   58&   21&   16&   47\cr 
   Q&      71&   78&   93&   58&   45&   40&   35&\qquad\qquad&  75&   60&   74&   56&   45&   38&   34\cr 
$\chi^2$&   4&   22&  100&   91&   46&   21&   27&\qquad\qquad&   2&    4&   42&   69&   43&   84&   35\cr 
   V&      73&   80&   95&   58&   47&   39&   34&\qquad\qquad&  75&   60&   74&   56&   45&   39&   34\cr 
$\chi^2$&  12&   19&  159&   46&  110&   33&   33&\qquad\qquad&   2&    8&   78&   34&  120&   38&   36\cr 
   W&      73&   80&   95&   57&   46&   39&   33&\qquad\qquad&  75&   59&   74&   57&   46&   40&   34\cr 
$\chi^2$&  22&   68&  104&  186&  111&   31&   18&\qquad\qquad&  11&   49&   58&   58&   99&   60&   47\cr 
\hline
\end{tabular}
\end{minipage}
\label{Tabl2}
\end{table*}

\section{Conclusion}

In this paper we have proposed  new methods of investigation
the statistical properties of the signal derived from the {\it WMAP}~
K--W bands foreground cleaned maps. With this approach we show
significant correlations between the phases
of the cleaned maps and the foregrounds. We have compared three cleaned
maps, ILC, TOH FCM and PCM, and have investigated variations of
the linear cross-correlation coefficients versus the multipole 
index, $\ell$. We suggest also more sophisticated ``friend--of--friend''
statistics. All the methods described in Sec. 3 show significant
cross-correlations between the phases of the cleaned maps and
the foregrounds which manifest themselves more clearly for the W
band of the {\it WMAP}. We have pointed out that some of
peculiarities of the ILC and PCM power spectrum, for
example, local minima at $\ell\sim 6$, and $\ell\sim 20-25$ are
accompanied by significant cross-correlations between phases 
of the foregrounds  and cleaned signal. Such correlations 
are important indicators for investigation and detection
of possible non-Gaussianity of the CMB signal which could 
be a byproduct of the component separation methods.

\section*{Acknowledgments}
This paper was supported by Danmarks Grundforskningsfond
through its support for the establishment of the Theoretical
Astrophysics Center. We thank Max Tegmark et al. for providing
their processed maps and making them public with openness.
We thank Igor Novikov and Lung-Yih Chiang for useful discussions.
We thank TAC CMB  collaboration for used GLESP code.
We also acknowledge the use of
HEALPix package
(G\'orski et al. 1999)\footnote{\tt http://www.eso.org/science/healpix/}
to produce $\alm$ from {\it WMAP} maps
and some figures.

\section{Appendix}

In this section we would like to introduce an analytical approach for
investigation
of the cross-correlation coefficient $r_{sf}(\ell)$ between foregrounds and
derived CMB
phases per each mode $\ell$. We define $r_{sf}(\ell)$ as
$$
r_{sf}^k (\ell)\equiv \frac{3}{4\pi^4l }\sum_{m=1}^\ell\int_{-\pi}^{\pi}
\int_{-\pi}^{\pi}d\Ph^{M} d\Psi^k\Ph^{M} \Psi^k
      \eqno (A1)
$$
where $k$ marks the foreground derived from the {\it WMAP} frequency band.
Here we assume that the distributions of both phases are close to homogeneous
ones with $\langle\Psi\rangle$=0, $\langle\Psi^2\rangle \approx \pi^2/3$
and we average a correlation coefficient over $m$ for a given $\ell$.
For the phase of the cleaned CMB signal we will use Eq.(\ref{eq7}), which
reflects directly the cross-correlation between the derived CMB
and foregrounds
phases. Then, using the definition $\Psi^j=\alpha_{jk} \Psi^k$,
where $\alpha_{jk}$
is the cross-correlation coefficient between the foreground phases
in $j$-th and
$k$-th channels, from Eq.(A1) we obtain
$$
r_{sf}^k (\ell)=-\frac{3}{\sqrt{2}\pi \ell}
   \sum_{m=1}^\ell\sum_j\frac{\w^{(j)}|\G^{(j)}|}{|\a^{cmb}|}
	  \frac{J_{\frac{3}{2}}(\pi \alpha_{jk})}{(\alpha_{jk})^{\frac{1}{2}}}
      \eqno (A2)
$$

Let us assume that cross-correlation between different foreground
phases is very strong and
corresponding coefficients are $\alpha_{ik}\approx1$\footnote{see Naselsky
et al. (2003) for details of the phase correlations.}.
Then, from (A2) we obtain
$$
r_{sf}^k (\ell)= -\frac{3}{\pi^2\ell}\sum_{m=1}^\ell \frac{1}{|\a^{cmb}|}\sum_j
\w^{(j)}|\G^{(j)}|
      \eqno (A3)
$$
For the PCM method, $\w^{(j)}$ coefficients are the functions of $\ell$,
but not $m$,
 while ratio $|\G^{(j)}|/|\a^{cmb}|$ depends on $m$ for a given value $\ell$.

From Eq.(\ref{eq5}) we get
$$
|\a^{M}-\a^{cmb}|=\left|\sum_j \w^{(j)}\G^{(j)}\right|
      \eqno (A4)
$$

and, because the foregrounds seem to be highly correlated,
$$
\left|\sum_j \w^{(j)}\G^{(j)}\right|=sign\left(\sum_j \w^{(j)}|\G^{(j)}|\right)
\sum_j\w^{(j)}|\G^{(j)}|,
$$

$$
r_{sf}^k (\ell)=-\frac{3}{\pi^2\ell}\sum_{m=1}^\ell\frac{|\a^{(M)}-\a^{cmb}|}
{|\a^{cmb}|}sign(|\a^{(M)}|-|\a^{cmb}|)~~(A5)
$$
Thus, the cross-correlation coefficient $r_{sf}^k (\ell)$ simply describes
the accuracy of the
reconstruction of the CMB signal.

\end{document}